\documentstyle[12pt]{article}

\begin{document}

\begin{flushright}
Imperial/TP/99-00/??\\ {\tt hep-ph/9912??}\\ {\LaTeX}-ed on
\today\\
\end{flushright}
\vspace{0.5in}
\begin{center}
{{\LARGE \bf Fluctuations and Phase Transition Dynamics
\footnote{To be published in the proceedings of the 4th Conference on
Quantum and stochastic gravity, String cosmology
   and Inflation, Peyresq, September 1999} }}\\ \baselineskip=18pt \vspace{1in}
{\large R.\ J.\ Rivers}\\ \vspace{.1in} {{\it Blackett Laboratory,
Imperial College, London SW7 2BZ}}\\

\end{center}

\vfill

\begin{abstract}
Kibble and Zurek have provided a unifying causal picture for the
appearance of classical defects like cosmic strings or vortices at the onset
of phase transitions in relativistic QFT and condensed matter
systems respectively.  In condensed matter the predictions
 are partially supported by agreement with
experiments in superfluid helium.   We provide an
alternative picture for the initial appearance of defects
that supports the experimental evidence. When the original
predictions fail, this is understood, in part, as a
consequence of thermal fluctuations (noise), which play a comparable role in both condensed matter and QFT.
\end{abstract}



\section{Overview}

In this talk I want to consider the emergence of 'classical' field
configurations - topological defects - after a phase transition, and
the extent to which thermal fluctuations can inhibit this
process.  This is of particular interest in the early universe, for
which we expect a sequence of transitions from a very symmetric
initial state, and in which the presence of classical defects can have
important astrophysical consequences.

The relevance of topological defects is that when symmetry breaks,
it does not do so uniformly.  At the very least, the field is
uncorrelated on the scale of the causal horizon at any time.  Since
a broken symmetry is, necessarily, characterised by degenerate
vacua, the choice of different vacua in domains in which the fields
are uncorrelated will lead naturally to topological defects between them as
the field does its best to order on large scales.  The nature of the
defects depends on the relevant homotopy group of the ground-state
manifold.
The most acceptable defect on cosmological grounds is the 'cosmic
string' - a generalised field vortex - which may have played in role
in structure formation.

Given a theory that permits vortices, at some time after the
transition we expect to find a network of them, behaving essentially
classically as Nambu-Goto strings, intersecting, chopping off loops
which decay, and straightening segments to reduce field gradients.
Simple calculations suggest late-time scaling solutions, with
similarity on a wide range of scales.

The details do not concern us here.  What interests us is how this
collection of essentially classical objects, which can be observed
directly, in principle, came into existence. The simplest
question, that we shall address here, is what is the density of
cosmic strings (or other defects) at the time of their appearance?

The early universe is very hot, but such a problem requires us to
go beyond equilibrium Thermal Field Theory. In practice, we often know remarkably
little about the dynamics of thermal systems.  For simplicity, I
shall assume scalar field  order parameters, with {\it continuous}
transitions. In principle, the field correlation length diverges
at a continuous transition.  In practice, it does not since there is not
enough time.   One possibility is that
the separation of 'defects' is characterised by the
correlation length when it checks its growth.  If this were simply so, a measurement of
defect densities would be a measurement of correlation lengths.
Estimates of this early field ordering and its contingent defects
in the early universe have
been made by Kibble\cite{kibble1,kibble2}, using simple
thermal\cite{kibble1} arguments or causal arguments\cite{kibble2}
different from the one above (although that is also due to
Kibble\cite{kibble3}).

There are great difficulties in converting such predictions for
the early universe into experimental observations since, but for a
 possible stray monopole, we have no direct evidence for them having existed
 \footnote{Although this does not impede our ability to make predictions for
 defect-driven fluctuations in the CMB, for example}. Zurek
suggested\cite{zurek1} that similar arguments to those in
\cite{kibble2} were applicable to
condensed matter systems for which direct experiments on defect
densities could be performed.  This has lead to considerable
activity from theorists working on the boundary between QFT and
Condensed Matter theory and from condensed matter
experimentalists. To date almost all experiments have involved
superfluids, for which vortices can be produced readily.
All but one experiment is in agreement with these simple
causal predictions and we shall pay particular attention to this
one failure of prediction.
In this talk I shall
\begin{itemize}
\item
review the Kibble/Zurek causality predictions for initial
correlation lengths and defect densities.
\item
summarise the results of the condensed matter experiments and
present an alternative picture for the onset of defect production
for condensed matter systems. I shall then show how this
alternative picture gives essentially the same results as the
Zurek picture for those condensed matter systems for which there
is experimental agreement.
\item
provide an explanation for why some condensed matter experiments
will be in disagreement with Zurek's predictions, including the
experiment in question.  We shall suggest that the prediction fails, in
part, because of the presence of thermal noise.
\item
use these ideas to address the more complicated problem of the
appearance of 'classical' defect configurations in QFT in the
light of Kibble's predictions, and the role of thermal noise in them.
\end{itemize}

\section{When symmetry breaks, how big are the smallest identifiable pieces?}

Defects in the large-scale ordering of the field can only appear
once the transition has taken place.
If it is the case that defect density can be identified simply from
the field
correlation length, the {\it maximum} density (an experimental
observable in condensed matter systems, although not for the early
universe) will be associated with the {\it smallest} identifiable
correlation length in the broken phase once the transition has
been effected. This provides the initial condition for the
evolution of field ordering. From this viewpoint, we can observe
the defects by default merely by determining the correlation
length at that time.

In order to see how to  identify these
'smallest pieces'\footnote{The
title of this section is essentially that posed in recent papers by
Zurek\cite{zurek2}.} it is sufficient to consider the simplest
theory with vortices, that of a
single relativistic {\it complex} scalar field in three spatial dimensions,
undergoing a
temperature quench.  In the first
instance we assume
that the qualitative dynamics
of the transition are conditioned by the field's
{\it equilibrium} free energy, of the form
\begin{equation}
F(T) = \int d^{3}x\,\,\bigg(|\nabla\phi|^{2}
+ m^{2}(T)|\phi|^{2} + \lambda |\phi|^{4}\bigg)
\label{FR2}
\end{equation}
Prior to the transition, at temperature $T>T_{c}$, the critical
temperature, $m (T) > 0$ plays the role of an effective 'plasma'
mass due to the interactions of $\phi$ with the heat bath, which
includes its own particles. After the transition, when $T$ is
effectively zero, $m^{2}(0) = - M^{2}< 0$ enforces the $U(1)$
symmetry-breaking, with field expectation values $\langle|\phi |
\rangle = \eta$, $\eta^{2} = M^{2}/\lambda$. The change in
temperature that leads to the change in the sign of $m^{2}$ is
most simply understood as a consequence of the system expanding.
Thus, in the early universe, a weakly interacting relativistic
plasma at temperature $T\gg M$ has an entropy density $s\propto
T^{3}$. As long as thermal equilibrium can be maintained, constant
entropy $S$ per comoving volume,  $S\propto s a(t)^{3}$, gives
$T\propto a(t)^{-1}$ and falling, for increasing scale factor
$a(t)$. Models that attempt to take
inflation into account, however, lead to 'preheating' that is not
Boltzmannian\cite{juan}.  Nonetheless, even in such cases it is
possible to isolate an effective temperature for long-wavelength
modes.  This is all that is necessary, but is too sophisticated
for the simple scenarios that we shall present here.  We shall not
even include a metric in Eq.\ref{FR2}.

The minima of the final
potential of  Eq.\ref{FR2} now constitute the circle $\phi = \eta
e^{i\alpha}$.
When the transition starts $\phi$ begins to fall into the valley
of the potential, choosing a random phase.
This randomly chosen
phase can vary from point to point subject to continuity.
At late times the failure of
the field to be uniform in phase on large scales will lead to it twisting
around classical 'defects' - solutions to $\delta F/\delta\phi =
0$ that locally minimise the energy stored in field gradients and
potentials. Those of interest to us are vortices, tubes of 'false' vacuum
$\phi\approx 0$, around which the field phase changes by $\pm
2\pi$. In an early universe context these are the simplest
possible 'cosmic strings'.

How this collapse takes place determines the size of the first
identifiable domains.  It was suggested by Kibble and Zurek
that this size is
essentially the equilibrium field correlation length $\xi_{eq}$ at some appropriate
temperature close to the transition.
I shall argue later that this is too simple but, nonetheless,
it is a plausible starting point.  Two very different
mechanisms have been
proposed for estimating this size.

\subsection{Thermal activation}

In the early work on the cosmic string scenario
an alternative possibility to simple causality  was to
assume\cite{kibble1}  that
initial domain size was fixed in the Ginzburg regime
by the correlation length at that time, rather than the causal radius.
By this we mean the following.
Once we are below $T_{c}$, and the central hump in
$V(\phi) =m^{2}(T)|\phi |^{2} + \lambda |\phi |^{4}$ is forming,
$T_{G}$ signals the temperature above which there is a
significant probability for thermal fluctuations over the central
hump on the scale of the correlation length.  Most simply, it is
determined by the condition
\begin{equation}
\Delta V(T_{G})\xi_{eq}^{3}(T_{G})\approx T_{G}
\end{equation}
where $\Delta V(T)$ is the difference between the central maximum
and the minima of $V(\phi ,T)$.  We find
$|1-T_{G}/T_{c}| = O(\lambda )$.

Whereas, above $T_{G}$ there will be a population of
'domains', fluctuating in and out of existence,
at temperatures below $ T_{G}$ fluctuations from one minimum to
the other become increasingly unlikely. When this happens the
correlation length is
\begin{equation}
\xi_{eq}(T_{G}) =
O\bigg(\frac{\xi_{0}}{\sqrt{1-T_{G}/T_{c}}}\bigg), \label{xiG}
\end{equation}
where $\xi_{0} = M^{-1}$ is the natural unit of length, the Compton
wavelength of the $\phi$ particles.

It is tempting\cite{kibble1,copeland} to identify $\xi_{eq}(T_{G})$ with the scale at
which stable domains begin to form. We shall see later that this
is incorrect, for quenches that are not too slow.
However, some care is needed if (as can happen in
condensed matter physics) we never leave the Ginzburg regime.

The formation of large domains is an issue that
requires more than equilibrium physics.  The most simple dynamical arguments
can be understood in terms of causality.

\subsection{Causality}

We have already mentioned that causality puts an {\it upper} bound
on domain size.  Specifically, if $G(r,t)$ is the two-field
correlation function at time $t$ for separation $r$, then $G$
vanishes for $r\geq 2t$ approximately. This was used by Kibble\cite{kibble3} to
put an upper bound on monopole density in the early universe.
If this causal bound and the Ginzburg criteria attempt to set scales once the critical
temperature has been {\it passed}, the causal arguments considered
now attempt to set scales {\it before} it is reached.

Here we attempt a {\it lower} bound on domain size, an upper bound
on defect density.
Suppose the temperature $T(t)$
varies sufficiently slowly with time $t$ that it makes sense to
replace $V(\phi ,T)$ by $V(\phi , T(t))$.
With $m^{2}(T(t))$ vanishing at $T=T_{c}$, which we suppose  happens at
$t=0$, the {\it equilibrium}
correlation length of the field fluctuations $\xi_{eq}(T(t))= |m^{-1}
(T(t))|$ diverges at $T(t) =T_{c}$.
It is sufficient to adopt a mean-field approximation in
which $m^{2}(T)\propto (T-T_{c})$.
The true correlation length $\xi (t)$ cannot diverge like $\xi_{eq}(T(t))$,
since it can only grow so far in a finite time.

Initially, for $t<0$, when we are far from the transition, we
again assume effective equilibrium, and the field correlation
length $\xi (t)$ tracks $\xi_{eq}(T(t))$ approximately. However,
as we get closer to the transition $\xi_{eq}(T(t))$ begins to
increase arbitrarily fast.  As a crude upper bound, the true
correlation length fails to keep up with $\xi_{eq}(T(t))$ by the
time $-{\bar t}$ at which $\xi_{eq}$ is growing at the speed of
light, $d\xi_{eq}(T(-{\bar t}))/dt =1$.  It was suggested by
Kibble\cite{kibble2} that, once we have reached this time $\xi
(t)$ {\it freezes} in, remaining approximately constant until the
time $t\approx +{\bar t}$ after the transition when is once again
becomes comparable to the now {\it decreasing} value of
$\xi_{eq}$. The correlation length
$\xi_{eq}({\bar t})=\xi_{eq}(-{\bar t})$ is argued to provide the scale
for the minimum domain size {\it after} the transition.

Specifically, if we assume a time-dependence $m^{2}(t) =
-M^{2}t/t_Q$ in the vicinity of $t=0$, when the transition begins to
be effected, then the causality condition gives
$t_C=t_{Q}^{1/3}(2M)^{-2/3}$.  As a result,
\begin{equation}
M\xi_{eq}({\bar t}) = (M\tau_{0})^{1/3},
\label{xiC0}
\end{equation}
which, with condensed matter in mind, we write as
\begin{equation}
{\bar\xi} = \xi_{eq}({\bar t}) = \xi_{0}\bigg(\frac{\tau_{Q}}{\tau_{0}}\bigg)^{1/3}
\label{xiC}
\end{equation}
where $\tau_{0}  = \xi_{0}= M^{-1}$ are the natural time and
distance scales.  In
contrast to  Eq.\ref{xiG},  Eq.\ref{xiC} depends explicitly on the
quench rate, as we would expect.

\subsection{QFT or Condensed Matter}

This approach of Kibble was one of the motivations for a similar
analysis by Zurek\cite{zurek1} of transitions
 in condensed matter. Qualitatively, neither the
Ginzburg thermal fluctuations, with fluctuation length
 Eq.\ref{xiG}, nor the simple causal argument above depend
critically on the fact that the free energy  Eq.\ref{FR2} is
originally assumed to be derived from a relativistic {\it quantum}
field theory.  After rescaling, $F$ could equally well be the
Ginzburg-Landau free energy for  the complex order-parameter field whose
magnitude determines the superfluid density. That is,
\begin{equation}
F(T) = \int d^{3}x\,\,\bigg(\frac{\hbar^{2}}{2m}|\nabla\phi |^{2}
+\alpha (T) |\phi |^{2} + \frac{1}{4}\beta |\phi |^{4}\bigg),
\label{FNR2}
\end{equation}
in which $\alpha (T)\propto m^{2}(T)$ vanishes at the critical
temperature $T_c$.  The only difference is that, in the causal
argument, the speed of light should be replaced by the speed of
(second) sound, with different critical index.

Explicitly, let us assume the mean-field result $\alpha (T) =
\alpha_{0}\epsilon (T )$, where $\epsilon = (T/T_c -1)$,
remains valid as $T/T_c$ varies with time $t$. In particular, we
first take  $\alpha (t)=\alpha (T(t))=-\alpha_{0} t/\tau_{Q}$ in
the vicinity of $T_c$.  The fundamental length scale $\xi_{0}$
is given from  Eq.\ref{FNR2} as
 $\xi_{0}^{2} = \hbar^{2}/2m\alpha_{0}$.  The Gross-Pitaevski theory suggests a
natural time-scale $\tau_{0} = \hbar /\alpha_{0}$. When, later, we
adopt the time-dependent Landau-Ginzburg (TDLG) theory we find
this still to be true, empirically, at order-of-magnitude level, and we keep
it.

 It follows that the equilibrium correlation length $\xi_{eq}
(t)$ and the relaxation time $\tau (t)$ diverge when $t$ vanishes
as
\begin{equation}
\xi_{eq} (t) = \xi_{0}\bigg|\frac{t}{\tau_{Q}}\bigg|^{-1/2},
\,\,\tau (t) = \tau_{0}\bigg|\frac{t}{\tau_{Q}}\bigg|^{-1}.
\end{equation}
The speed of sound is $c(t) =\xi_{eq} (t)/\tau (t)$, slowing down as
we approach the transition as $|t|^{1/2}$.  The causal counterpart
to $d\xi_{eq}
(t)/dt = 1$ for the relativistic field is
$d\xi_{eq}
(t)/dt = c(t)$.  This is satisfied at $t=-{\bar t}$, where ${\bar t}
=\sqrt{\tau_{Q}\tau_{0}}$, with corresponding correlation length
\begin{equation}
{\bar\xi} = \xi_{eq}({\bar t}) =\xi_{eq}(-{\bar t}) = \xi_{0}\bigg(\frac{\tau_{Q}}{\tau_{0}}\bigg)^{1/4}.
\label{xiZ}
\end{equation}
(cf.  Eq.\ref{xiC}).
A variant of this argument that gives essentially the same results
is obtained by comparing the quench rate directly to the relaxation rate of
the field fluctuations.  We stress that, yet again, the assumption is that the
length scale that determines the initial correlation length of the
field freezes in {\it before} the transition begins.

\section{Experiments}

The end result of the simple causality arguments is that, both for
QFT and condensed matter, when the
field begins to order itself its correlation length has the form
\begin{equation}
{\bar\xi} = \xi_{eq}({\bar t}) = \xi_{0}\bigg(\frac{\tau_{Q}}{\tau_{0}}\bigg)^{\gamma}.
\label{xiKZ}
\end{equation}
for appropriate $\gamma$.  \footnote{In fact, the powers of
 Eq.\ref{xiC} and  Eq.\ref{xiZ} are mean-field results, changed on
implementing the renormalisation group.}

The jump that Kibble made\cite{kibble2} in QFT was to assume that the correlation
 length  Eq.\ref{xiKZ}
 also sets the scale for the typical minimum intervortex
distance.
That is, the {\it initial} vortex density
$n_{def}$ is\footnote{Equivalently, the length of vortices in
a box volume $v$ is $O(n_{def}v)$.}
is assumed to be
\begin{equation}
n_{def} = \frac{1}{f^2}\frac{1}{{\bar\xi}^{2}}=
\frac{1}{f^2\xi_{0}^{2}}\bigg(\frac{\tau_{0}}{\tau_{Q}}\bigg)^{2\gamma}.
\label{ndef}
\end{equation}
for $\gamma = 1/3$ and $f=O(1)$.
We stress that this assumption is {\it independent} of the argument
that lead to  Eq.\ref{xiC}.
Since $\xi_{0}$ also measures cold vortex thickness, $\tau_{Q}\gg
\tau_{0}$ corresponds to a measurably large number of widely
separated vortices.

Even if cosmic strings were produced in so simple a way in the very
early universe it is not possible to compare the density
 Eq.\ref{ndef} with experiment, in large part because of our
uncertainty as to what is the appropriate theory.
  It was Zurek who
first suggested that this causal argument for defect density be
tested in condensed matter systems.

\subsection{Superfluid helium}

Vortex lines in both superfluid $^{4}He$ and $^{3}He$ are good analogues of global
cosmic strings. In $^{4}He$ the bose superfluid is
characterised by a complex field $\phi$, whose squared modulus
$|\phi |^{2}$ is the superfluid density.  The Landau-Ginzburg theory for $^{4}He$
has, as its free energy,
$F(T)$ of  Eq.\ref{FNR2}.
The static classical field equation $\delta F/\delta\phi = 0$ has
vortex solutions as before.  Specifically, a simple (winding number
unity) static straight vortex along the z-axis has
the form
\begin{equation}
\Phi ({\bf x}) = h(r)e^{\pm i\theta },
\label{vortex}
\end{equation}
where $\theta = \arctan (y/x)$ and $r^2 = x^2 + y^2$. For small
$r$, $h(r)= O(r)$, and for large $r$, $h(r) = \eta( 1-O(\xi_{0}^2
/r^2 ))$,
with effective width $\xi_{0}$.

The situation is more complicated, but more interesting, for
$^{3}He$. The reason is that $^{3}He$ is a {\it
fermion}.  Thus the mechanism for superfluidity is very different
from that of $^{4}He$.  Somewhat as in a BCS superconductor, these
fermions form the counterpart to Cooper pairs.  However, whereas the
(electron) Cooper pairs in a superconductor form a $^{1}S$ state,
the $^{3}He$ pairs form a $^{3}P$ state. The order parameter
$A_{\alpha i}$ is a complex $3\times 3$ matrix $A_{\alpha i}$.
There are two distinct superfluid phases, depending on how the
$SO(3)\times SO(3)\times U(1)$ symmetry is broken. If
the normal fluid is cooled at low pressures, it makes a transition
to the $^{3}He-B$ phase.

The Landau-Ginzburg free energy is, necessarily, more complicated,
permitting many types of vortex\cite{volovik2},
but the effective potential $V(A_{\alpha i}, T)$ has the diagonal
form\cite{Bunkov}
 $V(A, T) = \alpha
(T)|A_{ai}|^{2} + O(A^{4})$ for small fluctuations, and this is all
that we need for the production of vortices at very early times.
Thus the Zurek analysis leads to
the prediction  Eq.\ref{ndef}, as before, for appropriate $\gamma$.

\subsection{Counting vortices}

Although $^{3}He$ is more complicated to work with, the experiments to check
 Eq.\ref{ndef} are cleaner, since  even individual vortices can be detected by
magnetic resonance. Second, because vortex width is many atomic
spacings the Landau-Ginzburg theory is good ($\gamma = 1/4$).

So far, experiments have been of two types.  In the Helsinki
experiment\cite{helsinki} superfluid $^{3}He$ in a rotating
cryostat is bombarded by slow neutrons.  Each neutron entering the
chamber releases 760 keV, via the reaction $n + ^{3}He\rightarrow
p + ^{3}He + 760 keV$.  The energy goes into the kinetic energy of
the proton and triton, and is dissipated by ionisation, heating a
region of the sample above its transition temperature.  The heated
region then cools back through the transition temperature,
creating vortices. Vortices above a critical size  grow and migrate to the
centre of the apparatus, where they are counted by an NMR
absorption measurement.  The quench is very
fast, with $\tau_{Q}/\tau_{0} = O(10^{3})$.   Agreement with
 Eq.\ref{ndef} and  Eq.\ref{xiZ} is good, at the level of an order
of magnitude.  This is even though it is now argued\cite{kopnin}
that the Helsinki experiment should {\it not} show agreement
because of the geometry of the heating event.

 The second type of
experiment has been performed at Grenoble and
Lancaster\cite{grenoble}. Rather than count individual vortices,
the experiment detects the total energy going into vortex
formation. As before, $^{3}He$ is irradiated by neutrons. After
each absorption the energy released in the form of quasiparticles
is measured, and found to be less than the total 760 keV. This
missing energy is assumed to have been expended on vortex
production.  Again, agreement with Zurek's prediction
 Eq.\ref{ndef} and  Eq.\ref{xiZ} is good.

The experiments in $^{4}He$, conducted at Lancaster, follow
Zurek's original suggestion.  The idea is to expand a sample of
normal fluid helium so that it becomes superfluid at essentially
constant temperature. That is, we change $1-T/T_c$ from
negative to positive by reducing the pressure and increasing
$T_c$. As the system goes into the superfluid phase a tangle
of vortices is formed, because of the random distribution of field
phases.  The vortices are detected by scattering second sound off
them, and its attenuation gives a good measure of vortex
density.  A mechanical quench is slow, with $\tau_{Q}$ some tens
of milliseconds, and $\tau_{Q}/\tau_{0} = O(10^{10})$\footnote{For
$^{4}He$ mean-field theory is poor, and a better value for $\gamma$
is $\gamma = 1/3$.}.  Two
experiments have been performed\cite{lancaster,lancaster2}. In the
first fair agreement was found with the prediction  Eq.\ref{ndef},
but the second experiment has failed to see any vortices
whatsoever.

There is certainly no agreement, in this or any other experiment on $^{3}He$,
with the thermal fluctuation density that would be based on
 Eq.\ref{xiG}.

\section{The Kibble-Zurek picture for the value of ${\bar\xi}$ is correct.}

To do better than the simple causality arguments  we need a
concrete model for the dynamics.

\subsection{Condensed matter: the TDLG equation at early times}

We assume that the dynamics
of the transition can be derived from the explicitly time-dependent
Landau-Ginzburg free energy
\begin{equation}
F(t) = \int d^{3}x\,\,\bigg(\frac{\hbar^{2}}{2m}|\nabla\phi |^{2}
+\alpha (t)|\phi |^{2} + \frac{1}{4}\beta |\phi |^{4}\bigg).
\label{F}
\end{equation}
obtained from Eq.\ref{FNR2} on identifying $\alpha (t) = \alpha
(T(t)) = \alpha_{0}\epsilon (t)$, where $\epsilon = (T/T_c
-1)$.
In a quench in which $T_c$ or $T$ changes it is convenient to shift the origin in
time, to write $\epsilon (t)$  as
\begin{equation}
\epsilon (t) = \epsilon_{0} - \frac{t}{\tau_{Q}}\theta (t)
\label{eps}
\end{equation}
for $-\infty < t < \tau_{Q}(1 + \epsilon_{0})$, after which
$\epsilon (t) = -1$.  $\epsilon_{0}$
measures the original relative temperature and $\tau_{Q}$
defines the quench rate.  The quench begins at time $t = 0$ and the
transition from the normal to the superfluid phase begins at time $t
= \epsilon_{0}\tau_{Q}$.  Times subsequent to that are defined by
$\Delta t = t-t_0$.

Motivated by Zurek's later numerical\cite{zurek2} simulations, we
adopt the time-dependent Landau-Ginzburg (TDLG) equation for $F$,
on expressing $\phi$ as $\phi = (\phi_{1} + i\phi_{2})/\sqrt{2}$,
that
\begin{equation}
\frac{1}{\Gamma}\frac{\partial\phi_{a}}{\partial t} =
-\frac{\delta F}{\delta\phi_{a}} + \eta_{a}, \label{tdlg}
\end{equation}
where $\eta_{a}$ is Gaussian thermal noise, satisfying
\begin{equation}
\langle\eta_{a} ({\bf x},t)\eta_{b} ({\bf y}',t')\rangle =
2\delta_{ab}T(t)\Gamma\delta ({\bf x}-{\bf y})\delta (t -t').
\label{noise}
\end{equation}
This is a crude approximation for $^{4}He$, and a simplified form
of a realistic description of $^{3}He$ but it is not a useful
description of QFT, as it stands.

It is relatively simple to determine the
validity of Zurek's argument since it assumes that freezing in of
field fluctuations occurs just before symmetry breaking begins. At
that time the effective potential $V(\phi ,T)$ is still roughly
quadratic and we can see later that, for the relevant
time-interval $-{\bar t}\leq \Delta t\leq {\bar t}$ the
self-interaction term can be neglected ($\beta =0$).

In space, time and temperature units in which $\xi_{0} = \tau_{0} =
k_{B} =1$,  Eq.\ref{tdlg} then becomes
\begin{equation}
{\dot\phi}_{a}({\bf x},t) = - [-\nabla^{2} + \epsilon (t)]\phi_{a}
({\bf x},t) +{\bar\eta}_{a} ({\bf x},t). \label{free}
\end{equation}
where ${\bar\eta}$ is the renormalised noise. The solution of the
'free'-field linear equation is straightforward, giving a Gaussian
equal-time correlation function\cite{ray,ray3}
\begin{equation}
\langle\phi_{a} ({\bf r},t)\phi_{b} ({\bf 0},t)\rangle
=\delta_{ab}G({\bf r},t)
\label{diag}
\end{equation}
where
\begin{equation}
G(r, t) = \int_{0}^{\infty} d\tau\,{\bar T}(t-\tau/2)
\,\bigg(\frac{1}{4\pi\tau}\bigg)^{3/2} e^{-r^{2}/4\tau}\,e^{-\int_{0}^{\tau} ds\,\,\epsilon
(t- s/2)}.
\label{lgcorr}
\end{equation}
and ${\bar T}$ is the renormalised temperature.
At
time
$t_{0} = \epsilon_{0}\tau_{0}$, when the transition begins,  a
saddle-point calculation shows that, provided the quench is not
too fast,
\begin{equation}
G(r,t_{0})\approx \frac{T_c}{4\pi r}\,e^{-a(r/{\bar\xi})^{4/3}},
\label{notyuk}
\end{equation}
where $a = O(1)$, confirming Zurek's result.

Zurek's prediction is robust, since
further  calculation shows that
 $\xi (t)$ does not vary strongly in the interval
$-{\bar t} \leq \Delta t\leq {\bar t}$, where $\Delta t = t-t_{0}$.

\subsection{QFT: Closed time-path ensemble averaging at early times}

For QFT the situation is rather different. In the previous section,
instead of working with the TDLG equation, we
could have worked with the equivalent Fokker-Planck equation for the
probability $p^{FP}_{t}[\Phi ]$ that, at time $t>0$, the
measurement of $\phi$ will give the function $\Phi ({\bf x})$.
Thus $G(r,t)$ of  Eq.\ref{diag} can be written as
\begin{equation}
\delta_{ab}G({\bf r},t) = \langle\phi_{a} ({\bf r},t)\phi_{b} ({\bf 0},t)\rangle
=\int {\cal D}\Phi\,p^{FP}_{t}[\Phi ]\Phi_{a} ({\bf r})\Phi_{b} ({\bf
0}).
\label{corrFP}
\end{equation}

When solving the dynamical equations for a hot quantum field  it is convenient to work
with probabilities from the start.
Taking $t=0$ as our starting time for the evolution of the complex
field $\phi$  suppose that, at this time, the
system is in a pure state, in which the measurement of $\phi$ would
give $\Phi_0({\bf x})$. That is:-
\begin{equation}
\hat{\phi}(t=0,{\bf x}) | \Phi_0,t=0 \rangle = \Phi_0 | \Phi_0,t=0 \rangle.
\end{equation}
The probability $p_{t}[\Phi]$ that, at time $t>0$, the
measurement of $\phi$ will give the function $\Phi ({\bf x})$, is the double path integral
\begin{equation}
p_{t}[\Phi] = \int_{\phi_{\pm}(0) = \Phi_0}^{\phi_{\pm}(t) =
\Phi} {\cal D} \phi_+  {\cal D} \phi_-
\, \exp \biggl \{ i \biggl ( S_t [\phi_+]-S_t [\phi_-] \biggr ) \biggr \},
\end{equation}
where ${\cal D}\phi_{\pm} ={\cal D}\phi_{\pm ,1}{\cal D}\phi_{\pm
,2} )$
and $S_t [\phi]$ is the (time-dependent) action
obtained from Eq.\ref{FR2}, on substituting $m(t) = m(T(t))$ for $m(T)$.

$p_{t}[\Phi]$ can be
written in the closed time-path form in which,
 instead of separately integrating $\phi_{\pm}$ along the
time paths $0 \leq t \leq t_f$, the integral can be interpreted as
time-ordering of a field $\phi$ along the closed path $C_+ \oplus
C_-$ of Fig.1, where $\phi =\phi_+$ on $C_+$ and $\phi= \phi_-$ on $C_-$.
 When we extend
the contour from $t_f$ to $t= \infty$ either $\phi_+$ or $\phi_-$
is an equally good candidate for the physical field, but we choose
$\phi_+$.

The choice of a pure state at time $t=0$ is too simple to be of
any use. For simplicity, we assume that $\Phi$ is Boltzmann
distributed at time $t=0$ at an effective temperature of $T_0 =
\beta_0^{-1}$ according to the  Hamiltonian $H[\Phi]$
corresponding to the free-field action $S[\phi ]$, obtained by
setting $\lambda = 0$ in  Eq.\ref{FR2}, in which $\phi$ is taken
to be periodic in imaginary time with period $\beta_0$. We now
have the explicit form for $p_{t}[\Phi]$,
\begin{equation}
p_{t} [ \Phi] = \int_B {\cal D} \phi \, e^{i S_C [\phi]} \, \delta [
\phi_+ (t_f) - \Phi ],
\end{equation}
 written as the time ordering of a
single field along the contour $C=C_+ \oplus C_- \oplus C_3$, extended to include a
third imaginary leg, where $\phi$ takes the values $\phi_+$, $\phi_-$
and $\phi_3$ on $C_+$, $C_-$ and $C_3$ respectively, for which $S_C$
is $S[\phi_+]$, $S[\phi_-]$ and $S_0[\phi_3]$.
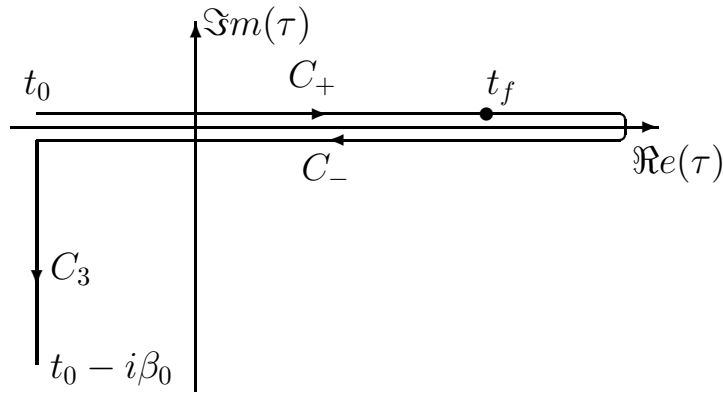
\begin{figure}[htb]
\begin{center}
\setlength{\unitlength}{0.5pt}
\begin{picture}(495,280)(35,480)
\put( 70,565){\makebox(0,0)[lb]{\large $C_3$}}
\put(260,640){\makebox(0,0)[lb]{\large $C_-$}}
\put(185,750){\makebox(0,0)[lb]{\large $\Im m(\tau)$}}
\put(250,710){\makebox(0,0)[lb]{\large $C_+$}}
\put(510,645){\makebox(0,0)[lb]{\large $\Re e (\tau)$}}
\put(400,705){\makebox(0,0)[lb]{\large $t_f$}}
\put( 50,705){\makebox(0,0)[lb]{\large $t_0$}}
\put(70,490){\makebox(0,0)[lb]{\large $t_0-i \beta_0$}}
\thicklines
\put(400,690){\circle*{10}}
\put( 40,680){\vector( 1, 0){490}}
\put( 60,690){\vector( 1, 0){220}}
\put(280,690){\line( 1, 0){220}}
\put(500,680){\oval(10,20)[r]}
\put(500,670){\vector(-1, 0){220}}
\put(280,670){\line(-1, 0){220}}
\put( 60,670){\vector( 0,-1){110}}
\put( 60,560){\line( 0,-1){ 60}}
\put(180,480){\vector( 0, 1){280}}
\end{picture}
\end{center}
\caption{The closed timepath contour $C_+ \oplus C_-$, with the Boltzmann imaginary leg}
\end{figure}

Just as we had no need to calculate $p^{FP}[\Phi ]_{t}$ explicitly in condensed matter
we can average in QFT without
having to calculate $p_t[\Phi]$ explicitly.  Specifically,
\begin{equation}
G_{ab}(r,t) = \langle\Phi_a ({\bf r})\Phi_b ({\bf
0})\rangle_{t} = \int {\cal D}\Phi\,p_{t}[\Phi ]\Phi_{a} ({\bf r})\Phi_{b} ({\bf
0})
\end{equation}
 is given by
\begin{equation}
G_{ab}(r,t) = \langle\phi_a ({\bf r},t)\phi_b
({\bf 0},t)\rangle,
\label{wight}
\end{equation}
the equal-time thermal Wightman function with the given thermal boundary conditions.

Fortunately, as for the condensed matter case, the interval
$-{\bar t} \leq \Delta t\leq {\bar t}$ occurs in the {\it linear} regime,
when the self-interactions are unimportant. The relevant equation
for constructing the correlation functions of this one-field system
is now the second-order equation
\begin{equation}
\frac{\partial^{2}\phi_a}{\partial t^{2}} = -\frac{\delta
F}{\delta\phi_a},
\label{op}
\end{equation}
for $F$ of  Eq.\ref{FR2}. This is solvable in terms of the mode functions
$\chi^{\pm}_{k}(t)$, identical for $a=1,2$, satisfying
\begin{equation}
\Biggl [ \frac{d^2}{dt^2} + {\bf k}^2 + m^2(t) \Biggr ]\chi^{\pm}_{k}(t)  =0,
\label{mode}
\end{equation}
subject to $\chi^{\pm}_{k}(t)
=
e^{\pm i\omega_{in}t}$ at
$t\leq 0$,
for incident frequency $\omega_{in} = \sqrt{{\bf
k}^{2} +\epsilon_{0} M^{2}}$ and $m^2(t) = \epsilon (t)M^{2}$,
where $\epsilon (t)$ is parameterised as for the  TDLG equation
above.  This corresponds to a temperature quench from an initial
state of thermal equilibrium at temperature $T_{0}>T_c $, where
$(T_{0}/T_c -1) = \epsilon_{0}$.
The diagonal correlation function $G(r, t)$ of  Eq.\ref{diag} is given as the
equal-time propagator
\begin{eqnarray}
G(r, t)&=&\int d \! \! \! / ^3 k
\, e^{i {\bf k} . {\bf x} } \chi^{+}_{k}(t) \chi^{-}_{k}(t)C(k)
\\
\nonumber
&=& \frac{1}{2\pi^{2}}\int dk\,k^{2}\frac{\sin kr}{kr}
\chi^{+}_{k}(t) \chi^{-}_{k}(t)C(k),
\label{modesum}
\end{eqnarray}
where $ C(k) =\coth(\omega_{in} (k)/2T_0
)/2\omega_{in}(k)$ encodes the initial conditions.

An exact solution can be given\cite{bowick} in terms of Airy
functions. Dimensional analysis shows that, on ignoring the
k-dependence of $C(k)$, appropriate for large $r$ (or small $k$),
$\xi_{eq}({\bar t})$ of  Eq.\ref{xiC} again sets the scale of the
equal-time correlation function. Specifically,
\begin{equation}
G(r,t_0)\propto\int d\kappa\,\frac{\sin\kappa
(r/{\bar\xi})}{\kappa (r/{\bar\xi})}\,F(\kappa ),
\label{GQFT}
\end{equation}
where $F(0) = 1$ and $F(\kappa )\sim \kappa^{-3}$ for large $\kappa$.
 Kibble's insight is
correct, at least for this case of a single (uncoupled) field.

\section{Vortex densities do not determine correlation lengths directly}

We have seen that there is no reason to disbelieve the causal
arguments of Kibble for QFT and Zurek for condensed matter as to
the field correlation length at the time of the transition.
The excellent agreement with the $^{3}He$
experiments also shows that, despite the very interesting
simulations of Kopnin et al.\cite{kopnin}, this length does, indeed,
characterise vortex separation for condensed matter at the time when the defects form.

However, the recent Lancaster experiment shows that this
cannot always be the case. Significantly, for $^{3}He$ the
Ginzburg regime is extremely narrow, whereas for $^{4}He$ it is very
broad.  In fact, the $^{4}He$ experiments begin and end in the Ginzburg
regime, where thermal fluctuations dominate.  The causality
arguments are too simple to accommodate these facts.

If these differences are to be visible in the formalism, it can
only be through the way in which we relate  vortex density
to correlation length.  We have already observed that the TDLG
equation can be recast as the Fokker-Planck equation, whereby the
ensemble averages can be understood as averaging with respect to the
probability $p_{t}[\Phi ({\bf x})]$ that, at time $t$, the field takes
value $\Phi ({\bf x})$. We can use these probabilities, implicit in the
correlation functions, to estimate defect densities.

\subsection{Classical defects in condensed matter}

It would be foolish to estimate the probability of finding profiles like
$\Phi (\bf x)$ of Eq.\ref{vortex} directly.
One way is to work through {\it line zeroes}, since vortices have line zeroes of the complex
field $\phi$ at the centre of their cores.
The converse is not true since zeroes occur on
all scales.  However, a starting-point for counting
vortices in superfluids is to count
line zeroes of an appropriately coarse-grained field, in which
structure on a scale smaller than $\xi_{0}$, the classical vortex size, is
not present\cite{popov}.  That is, we do not want to entertain
vortices within vortices.
For the moment, we put in a cutoff $l = O(\xi_{0})$ by hand into the Fourier transform $G(k,t)$
of $G(r,t)$, as
\begin{equation}
G_{l}(r, t) = \int d \! \! \! / ^3 k\, e^{i{\bf k}.{\bf r}}G(k, t)\,e^{-k^{2}l^{2}}.
\label{Gl}
\end{equation}
We stress that  the {\it long-distance} correlation length
$\xi_{eq}({\bar t})$  depends essentially on the position of the
nearest singularity of $G(k, t)$ in the complex k-plane, {\it
independent} of $l$.

This is not the case for the line-zero density $n_{zero}$.
For example, in our Gaussian approximation of the previous section
 $n_{zero}$ can be calculated exactly from the two-point correlation
function $G(r,t)$ with $p_{t}[\Phi ]$ implicit.  It can be shown,
quite easily\cite{halperin,maz} that it depends on
the {\it short-distance} behaviour of $G_{l}(r,t)$ as
\begin{equation}
n_{zero}(t) = \frac{-1}{2\pi}\frac{G_{l}''(0,t)}{G_{l}(0, t )},
\label{nzero}
\end{equation}
the ratio of fourth to second moments of $G(k,t)\,e^{-k^{2}l^{2}}$.

There are several prerequisites before line zeroes can be identified
with vortex cores, and $n_{zero}(t)$ with $n_{def}(t)$.
\begin{itemize}
\item
The field, on average must have achieved its symmetry-broken
ground-state equilibrium value
\begin{equation}
\langle |\phi |^{2}\rangle = \alpha_{0}/\beta.
\label{eq}
\end{equation}
This, in itself, is sufficient to show that the causal time ${\bar
t}$ is {\it not} the time to begin looking for defects, since
$\langle |\phi |^{2}\rangle$ is small at this time.  This, in turn,
requires that $G(k,t)$ be non-perturbatively (in $\beta$) large.
\item
Only when  $\partial n_{zero}/\partial l$ is
small in comparison to
$n_{zero}/l$ at $l = \xi_{0}$ will the line-zeroes have the
non-fractal nature of classical defects on small-scales, although vortices
 may behave like random walks on larger scales.  As the power in the long
wavelength modes increases the 'Bragg'
peak develops in $k^{2}G(k,t)$, moving in towards $k=0$.  This
condition then becomes the condition that the peak dominates its tail.
\item
The energy in field gradients should be commensurate with the
energy in classical vortices with same density as that of line
zeroes.
\end{itemize}

We stress that these are necessary, but not sufficient, conditions
for classical vortices.  In particular, although they can be
satisfied in the self-consistent linear approximation that will be
outlined below, only the full nonlinearity of the system can
establish classical profiles.  We will term such zeroes as satisfy
these conditions proto-vortices. It has to be said that most (but
not all\cite{gleiser,calzetta}) numerical lattice simulations cannot
distinguish between proto-vortices and classical vortices.

\subsection{TDLG condensed matter}

We begin
with condensed matter, which we will find to be easier.
As the system evolves away from the transition time, the free
equation Eq.\ref{free} ceases to be valid, to be replaced by the full
equation
\begin{equation}
{\dot\phi}_{a}({\bf x},t) = - [-\nabla^{2} + \epsilon
(t)+{\bar\beta}|\phi ({\bf x},t)|^{2}]\phi_{a} ({\bf x},t)
+{\bar\eta}_{a} ({\bf x},t),
\label{full}
\end{equation}
where ${\bar\beta}$ is the rescaled coupling.

In order to retain some analytic understanding of the way that
the density is such an ideal quantity to make predictions for,
we  adopt the approximation of preserving Gaussian
fluctuations by linearising the self-interaction as
\begin{equation}
{\dot\phi}_{a}({\bf x},t) = - [-\nabla^{2} + \epsilon_{eff}
(t)]\phi_{a} ({\bf x},t) +{\bar\eta}_{a} ({\bf x},t),
\end{equation}
where $\epsilon_{eff} $ contains a (self-consistent) term
$O({\bar\beta}\langle |\phi |^{2}\rangle)$.
Additive renormalisation is necessary, so
that $\epsilon_{eff}\approx \epsilon$, as given earlier, for
$t\leq t_{0}$.

Self-consistent linearisation is the standard approximation in
non-equilibrium QFT\cite{boyanovsky,LA}, but is not strictly
necessary here, since numerical simulations that identify line
zeroes of the field can be made that use the full
self-interaction\cite{zurek2}. However, there are none that address
our particular problems exactly.  Given the similarities with the
QFT case, for which it is difficult to do much better than a Gaussian,
there are
virtues in comparing the Gaussian approximation for the two cases.

 The solution for $G(r,t)$ is a straightforward generalisation of
Eq.\ref{lgcorr},
\begin{equation}
G(r,t) =\int_{0}^{\infty} d\tau \,{\bar T}(t-\tau /2)
\bigg(\frac{1}{4\pi\tau}\bigg)^{3/2}
e^{-r^{2}/4\tau}\,e^{-\int_{0}^{\tau} ds\,\,\epsilon_{eff} (t-
s/2)},
\end{equation}
where  ${\bar T}$ is the rescaled temperature, as before.

Assuming a {\it single} zero of $\epsilon_{eff} (t)$ at $t =
t_{0}$, at $r=0$ the exponential in the integrand peaks at $\tau
={\bar\tau} = 2(t-t_{0})$, the counterpart of the Bragg peak in proper-time.
Expanding about ${\bar\tau}$ to
quadratic order gives\cite{ray3}
\begin{equation}
G_{l}(0,t)\approx {\bar
T}_{c}\,e^{2\int_{t_{0}}^{t}du\,|\epsilon_{eff} (u)|
}\int_{0}^{\infty} \frac{ d\tau\,e^{-(\tau -
2(t-t_{0}))^{2}|\epsilon '(t_{0})|/4}}{[4\pi (\tau + {\bar
l}^{2})]^{3/2} },
\label{Gsmall}
\end{equation}
where we have put in the momentum cutoff $k^{-1}> l =\bar{l}\xi_{0}
=O(\xi_{0} )$of Eq.\ref{Gl} by hand. For times $t >
\epsilon_{0}\tau_{Q}$ we see that, as the unfreezing occurs, long
wavelength modes with $k^{2} < t/\tau_Q - \epsilon_{0}$ grow
exponentially.

The effect of the back-reaction is to stop the growth of
$G_{l}(0,t)-G_{l}(0,t_{0})= \langle |\phi |^{2}\rangle_{t}-\langle
|\phi |^{2}\rangle_{0}$ at its symmetry-broken value
${\bar\beta}^{-1}$ in our dimensionless units.
A necessary condition for this is
$\lim_{u\rightarrow\infty}\epsilon_{eff} (u) = 0$. That is, we
must choose
$\epsilon_{eff}(t) = \epsilon (t)  +
{\bar\beta}(G_{l}(0,t)-G_{l}(0,t_{0}))$,
thereby preserving Goldstone's theorem.

At $t=t_{0}$, when the approximation Eq.\ref{Gsmall} is good,
both numerator and denominator are dominated by the
short wavelength fluctuations at small $\tau$. Even though the
field is correlated over a distance ${\bar\xi}\gg l$ the density
of line zeroes $n_{zero} = O(l^{-2})$ depends entirely on the
scale at which we look. In no way would we wish to identify these
line zeroes with prototype vortices. However, as time passes the
peak of the exponential grows and $n_{zero}$ becomes increasingly
insensitive to $l$. How much time we have depends on the magnitude
of ${\bar\beta}$, since once $G(0,t)$ has reached this value it
stops growing. The time $t^{*}$ at which this happens can be estimated by
substituting $\epsilon (u)$ for $\epsilon_{eff} (u)$ in the
expression for $G_{l}(0,t)$ above.

For $t>t^{*}$ the equation for $n_{zero}(t)$ is not so simple
since the estimate for $G_{l}(0,t)$ above, based on a single isolated zero of
$\epsilon_{eff} (t)$, breaks down because of the approximate
vanishing of $\epsilon_{eff} (t)$ for $t>t^{*}$.  A more careful
analysis shows that $G_{l}(0,t)$ can be written as
\begin{equation}
G_{l}(0,t)\approx \int_{0}^{\infty} \frac{ d\tau\,{\bar T}(t-\tau
/2)}{[4\pi (\tau + {\bar l}^{2})]^{3/2} } {\bar G}(\tau,t),
\end{equation}
where ${\bar G}(\tau,t)$ has the same peak as before at $\tau =
2(t-t_{0})$, in magnitude and position,
but ${\bar G}(\tau,t)\cong 1$ for $\tau < 2(t-t^{*})$. Thus, for
$\tau_{Q}\gg\tau_{0}$, $G_{l}(0,t)$ can be approximately separated
as
$G_{l}(0,t)\cong G_{l}^{UV}(t) + G^{IR}(t)$,
where
\begin{equation}
G_{l}^{UV}(t)= {\bar T}(t)\,\int_{0}^{\infty}
d\tau\,/[4\pi (\tau + {\bar l}^{2})]^{3/2},
\end{equation}
the counterpart of the Bragg 'tail', describes  the
scale-{\it dependent} short wavelength thermal noise, proportional
to temperature, and
\begin{equation}
G^{IR}(t) =\frac{{\bar T}_{c}}{(8\pi(t-t_{0}))^{3/2}}
\,\int_{-\infty}^{\infty}d\tau {\bar G}(\tau,t)
\end{equation}
describes the scale-{\it independent}, temperature independent,
long wavelength fluctuations. A similar decomposition
$G\prime\prime_{l}(0,t)\cong G\prime\prime_{l}^{UV}(t) +
G\prime\prime^{IR}(t)$ can be performed.  In particular,
$G\prime\prime^{IR}(t)/G^{IR}(t)= O(t^{-1})$.

Firstly, suppose that, for $t\geq t^*$,
  $ G^{IR}(t)\gg G_{l}^{UV}(t)$ and  $
G\prime\prime^{IR}(t)\gg G\prime\prime_{l}^{UV}(t)$, as would be
the case for a temperature quench ${\bar T}(t)\rightarrow 0$.
 Then, with little thermal noise, we have widely separated
line zeroes, with density $n_{zero}(t)\approx
-G\prime\prime^{IR}(t)/2\pi G^{IR}(t)$ and $\partial
n_{zero}/\partial l$ is small in comparison to $n_{zero}/l$ at $l =
\xi_{0}$.
Further, once the line zeroes have straightened on small scales at
$t>t^*$, the Gaussian field energy, largely in field gradients, is
\begin{equation}
{\bar F}\approx\langle\int_{V}
d^{3}x\,\frac{1}{2}(\nabla\phi_{a})^{2}\rangle= -VG''(0,t),
\end{equation}
where $V$ is the spatial volume. This matches the
energy
\begin{equation}
{\bar E}\approx V n_{def}(t)(2\pi G(0,t)) = -VG''(0,t)
\end{equation}
possessed by a network of classical global strings with density
$n_{zero}$, in the same approximation of cutting off their
logarithmic tails.

From our comments above, we identify such essentially non-fractal line-zeroes with
prototype vortices, and $n_{zero}$ with $n_{def}$.  Of course, we require
non-Gaussianity to create true classical energy profiles.
Nonetheless, the Halperin-Mazenko result may be well approximated
for a while even when the fluctuations are no longer
Gaussian\cite{calzetta}.

 For times $t>t^{*}$
\begin{equation}
n_{zero}(t)\approx \frac{{\bar t}}{8\pi (t-t_{0})
}\frac{1}{\xi_{0}^{2}}\sqrt{\frac{\tau_{0}}{\tau_{Q}}},
\end{equation}
the solution to Vinen's equation\cite{vinen}
\begin{equation}
\frac{\partial n_{zero}}{\partial t} = -\chi_{2}\frac{\hbar}{m}
n_{def}^{2},
\end{equation}
where $\chi_{2} = 4\pi$ in our approximation\footnote{Calculations for
$\chi_2$ for realistic values of $\xi_0$ and $\tau_0$ give $\chi_2 > 4\pi$
for both $^{4}He$ and $^{3}He$}.
What is remarkable in this approximation is that the
density of line zeroes uses {\it no} property of the self-mass
contribution to $\epsilon_{eff}(t)$, self-consistent or otherwise.

This decay law is assumed in the analysis of the Lancaster
experiments.
The empirical value of $\chi_2$ used in them
 is not taken from quenches, but turbulent
flow experiments. It is suggested\cite{lancaster2} that $\chi_{2}
\approx 0.005$, a good three orders of magnitude smaller than our
prediction above. Although the TDLG theory is not very reliable
for $^{4}He$, if our estimate is sensible it does imply that
vortices produced in a {\it temperature} quench decay much faster
than those produced in turbulence.

Equally importantly, we shall see that, for early
time at least, thermal fluctuations are large in the Lancaster
experiments. However, for $^{3}He$, with negligable UV
contributions, we estimate the primordial density of vortices as
\begin{equation}
n_{zero}(t^{*})\approx \frac{{\bar t}}{8\pi (t^{*} - t_{0})
}\frac{1}{\xi_{0}^{2}}\sqrt{\frac{\tau_{0}}{\tau_{Q}}},
\label{primdef}
\end{equation}
in accord with the original prediction of Zurek. Because of the
rapid growth of $G(0,t)$, $(t^{*}-t_{0})/{\bar t} =p > 1 = O(1)$.
We note that the factor\footnote{An errant factor of 3 appeared
 in the result of
\cite{ray}} of $f^{2}=8\pi p$ gives a value of $f = O(10)$, in
agreement with the empirical results of \cite{grenoble} and the
numerical results of \cite{zurek3}\footnote{The temperature
quench of the latter is somewhat different from that considered
here, but should still give the same results in this case}.

Whereas Eq.\ref{primdef} is appropriate for $^{3}He$, the
situation for the Lancaster $^{4}He$ experiments is complex,
since they are {\it pressure} quenches for which the temperature
$T$ is almost {\it constant} at $T\approx T_{c}$.
 Unlike
temperature quenches\cite{zurek2,boyanovsky2}, thermal
fluctuations here remain at full strength\footnote{Even for $^3
He$, $T/T_{c}$ never gets very small, and henceforth we take
$T=T_{c}$ in $G_{l}(0,t)$ above}. The necessary time-{\it
independence} of $ G^{IR}(t)$ for $t>t^*$ is achieved by taking
$\epsilon_{eff} (u)= O(u^{-1})$. In consequence, as $t$ increases
beyond $t^{*}$ the relative magnitude of the UV and IR
contributions to $G_{l}(0,t)$ remains {\it approximately
constant} at its value at $t = t^*$.

Nonetheless, as long as the UV fluctuations are insignificant at
$t=t^*$ the density of line zeroes will remain largely independent
of scale. This follows if $ G\prime\prime^{IR}(t^*)\gg
G\prime\prime_{l}^{UV}(t^*)$, since $G\prime\prime_{l}(0,t)$
becomes scale-independent later than $G_{l}(0,t)$. In \cite{ray}
we showed that this is true provided
\begin{equation}
(\tau_{Q}/\tau_{0})(1-T_{G}/T_{c})<C\pi^{4},
\label{ginlim}
\end{equation}
where $C= O(1)$ and $T_G$ is the Ginzburg temperature. With
$\tau_{Q}/\tau_{0} = O(10^{3})$ and $(1-T_G /T_c ) = O(10^{-12})$
this inequality is well satisfied for a linearised TDLG theory for
$^{3}He$ derived\footnote{Ignoring the position-dependent
temperature of \cite{kopnin}} from the full TDGL
theory\cite{Bunkov}, but there is no way that it can be satisfied
for $^{4}He$, when subjected to a slow mechanical quench, as in
the Lancaster experiment, for which $\tau_{Q}/\tau_{0} =
O(10^{10})$, since the Ginzburg regime is so large that $(1-T_G
/T_c ) = O(1)$. As far as the left hand side of Eq.\ref{ginlim} is
concerned, the $^{4}He$ quench is {\it nineteen}
orders of magnitude slower than its $^{3}He$ counterpart.

When the inequality is badly violated, as with $^{4}He$ for slow
pressure  quenches, then the density of zeroes $n_{def}=
O(l^{-2})$ after $t^*$ again depends explicitly on the scale $l$ at which
we look and they are not candidates for vortices. Since the whole
of the quench takes place within the Ginzburg regime this is not
implausible. However, it is possible that, even though the thermal
noise never switches off, there is no more than a postponement of
vortex production, since our approximations must break down at
some stage. The best outcome is to assume that the effect of the
thermal fluctuations on fractal behaviour is diminished, only
leading to a delay in the time at which vortices finally appear.
Even if we suppose that $n_{def}$ above is a starting point for
calculating the density at later times, albeit with a different
$t_{0}$, thereby preserving Vinen's law, we then have the earlier
problem of the large $\chi_{2} = O(f^2 )$, which would make it
almost
impossible to see vortices.

For all that, a numerical simulation that goes beyond the Gaussian
approximation specifically tailored to the Lancaster parameters is
crucial if we are to understand what is really happening. We hope
to pursue this elsewhere.

\section{The Appearance of Structure in QFT}

When, in Section 5.2 we set up the closed time-path formalism for
the field probabilities $p_{t}[\Phi ]$, our aim was the limited one
of establishing the role of Kibble's causal correlation length
${\bar\xi}$ in Eq.\ref{GQFT}. We now appreciate, from condensed matter
theory, that this does not, of  itself, imply vortices at that separation.

\subsection{Proto-vortices in QFT}

To establish a link between the correlation function $G(r,t)$ and
vortices is even more problematic in QFT than for condensed matter systems.
Yet again, we attempt to count vortices by counting line zeroes\cite{al}.
In the Gaussian approximations that we shall continue to adopt the
expression Eq.\ref{nzero} for $n_{zero}$ is equally applicable to
QFT.
This counting of zeroes is the
basis of numerous numerical simulations\cite{tanmay,andy,scherrer} of cosmic
string
networks built from Gaussian fluctuations.

The prerequisites for line zeroes in condensed matter that we posed
after  Eq.\ref{nzero} still stand for QFT (except that
$\langle|\phi^{2}|\rangle = M^{2}/\lambda$), but there are  further
complications peculiar to QFT.
In particular, in QFT we need to consider the whole density matrix $\langle\Phi '
|\rho (t)|\Phi \rangle$ rather than just the diagonal elements
$p_{t}[\Phi ] = \langle\Phi |\rho (t)|\Phi \rangle$.  Classicality is
understood in terms of 'decoherence', manifest most simply by the
approximate diagonalisation of the reduced density matrix on coarse-graining.
By this we mean the separation of the whole into the 'system', and
its 'environment' whose degrees of freedom are integrated over, to
give a reduced density matrix.  The environment can be either other
fields with which our scalar is interacting or even the short
wavelength modes of the scalar field itself \cite{muller2,lombardo}.
When interactions are taken into account this leads to quantum noise
and dissipation.

In the
Gaussian approximations that we shall adopt here, with
$\langle\Phi \rangle = 0$, integrating out short wavelengths with
$k>l^{-1}$ is just equivalent to a momentum cut-off at the same
value. This gives neither noise nor dissipation and diagonalisation does not occur.
Nonetheless, from our viewpoint of counting line-zeroes,
fluctuations are still present when $l = O(M^{-1})$ that can prevent us
from identifying line-zeroes with proto-vortices, if the quenches
are too slow.

For all these caveats, there are other symptoms of classical behaviour
once $G_{l}(0;t)$ is non-perturbatively large. Instead
of a field basis, we can work in a particle basis and measure the
particle production as the transition proceeds.
The
presence of a non-perturbatively large peak in $k^2 G(k;t)$ at $k =
k_0$ signals
a non-perturbatively large occupation number $N_{k_{0}}\propto 1/\lambda$
of particles at the same wavenumber $k_0$\cite{boyanovsky}.
With $n_{zero}$ of (\ref{nzero}) of order $k_{0}^{2}$
this shows that the long wavelength modes can now begin to be treated classically.
From a slightly different viewpoint, the Wigner functional only
peaks about the classical phase-space trajectory once the power is
non-perturbatively large\cite{guth,muller}.
More crudely, the diagonal density matrix
elements are only then significantly non-zero for
non-perturbatively large field configurations
$\phi\propto\lambda^{-1/2}$, like vortices.

\subsection{Mode growth v. fluctuations}

For early times we
revert to the mode decomposition of  Eq.\ref{mode}.  The term $coth
(\omega_{in}/2T_0 )$ appearing in it can be approximated by
$2T_{0}/\sqrt{\epsilon_{0}}M$.  Even though this is a temperature
quench, it shows strong similarities to the pressure quench of
condensed matter, since
both the long and short wavelength contributions to $G(r,t)$ are
scaled by the same temperature and we cannot switch off the latter.

The field
becomes ordered, as before, because of the exponential growth of
long-wavelength modes, which stop growing once the field has sampled
the groundstates.  What matters is the relative weight of these
modes (the 'Bragg' peak)
to the fluctuating short wavelength modes in the decomposition
 Eq.\ref{modesum} at this time, since the contribution of these latter is very
sensitive to the cutoff $l$.  Only if their contribution to
 Eq.\ref{ndef} is small when field growth stops can a network of
line-zeroes be well-defined at
early times, let alone have the predicted density.
Since the peak is non-perturbatively large this
requires small coupling, which we assume.

Consider a quench with $\epsilon
(t)$ as in  Eq.\ref{eps},
in which the symmetry-breaking begins at relative time $\Delta t =
t-t_0 = 0$.
For a {\it
free} roll,
the exponentially growing modes that appear when
$\Delta t>t^{-}_{k} = t_{Q}k^{2}/M^{2}$ lead to the approximate WKB solution\cite{ray2}
\begin{equation}
G(r;\Delta t)\propto
\frac{T}{M|m(\Delta t)|}\bigg(\frac{M}{\sqrt{\Delta
tt_Q}}\bigg)^{3/2}e^{\frac{4M\Delta t^{3/2}}
{3\sqrt{t_Q}}}
e^{-r^{2}/\xi^{2}(\Delta t)}
\label{Wexp}
\end{equation}
where
$\xi^{2}(\Delta t) = 2\sqrt{\Delta tt_Q}/M$.
The provisional freeze-in time $ t_*$ when
$\langle|\phi^{2}|\rangle = M^{2}/\lambda$
is then,  for
$Mt_{Q} < (1/\lambda)$,
\begin{equation}
M\Delta t_{*} \simeq (Mt_{Q})^{1/3}(\ln (1/\lambda ))^{2/3}
\simeq M{\bar t}(\ln (1/\lambda ))^{2/3},
\label{tfs}
\end{equation}
where $\Delta t_* = t_* - t_0$.
This is greater than $M{\bar t}$, but not by a large multiple.
Comparison with condensed matter, for which the ratio is a few
($3-5$) suggests that  we don't need a superweak theory\cite{ray2}.

At this qualitative level the correlation length at $t_*$ is
\begin{equation}
M^{2}\xi^{2}(t_{*})\simeq (Mt_{Q})^{2/3}(\ln (1/\lambda ))^{1/3}.
\label{chiss2}
\end{equation}
The effect of the other modes is larger than for the instantaneous
quench,  giving, at $t=t_{*}$
\begin{equation}
n_{zero}=  \frac{M^{2}}{\pi (M\tau_{Q})^{2/3}}
(\ln (1/\lambda ))^{-1/3}[1 + E].
\label{nisMf}
\end{equation}
The error term $E =O(\lambda^{1/2}(Mt_{Q})^{4/3}(\ln
(1/\lambda))^{-1/3})$ is due to oscillatory modes, sensitive to the cutoff.
In mimicry of  Eq.\ref{ndef} it is helpful to rewrite
 Eq.\ref{nisMf} as
\begin{equation}
n_{zero}=  \bigg[\frac{1}{\pi \xi_{0}^{2}}\bigg(\frac{\tau_{0}}{\tau_{Q}}\bigg)^{2/3}\bigg]
(\ln (1/\lambda ))^{-1/3}[1 + E].
\label{nisMf2}
\end{equation}
in terms of the scales $\tau_{0} = \xi_{0} = M^{-1}$.
The first term in  Eq.\ref{nisMf2} is the Kibble estimate of
 Eq.\ref{ndef}, the second is the small multiplying factor,
that yet again shows that estimate can be
correct, but for completely different reasons.  The third term shows
when it can be correct, since $E$ is also a measure of the
sensitivity of $n_{def}$ to the scale at which it is measured.
  The condition $E^{2}\ll 1$,
necessary for a proto-vortex network to be defined, is then guaranteed if
\begin{equation}
(\tau_{Q}/\tau_{0})^{2}(1-T_{G}/T_c )<C,
\label{ginlim2}
\end{equation}
$C = O(1)$, on using the relation $(1-T_{G}/T_c ) = O(\lambda )$.  This is
the QFT counterpart to  Eq.\ref{ginlim}.

For example, suppose that this approach is relevant to the local strings
of a strong Type-II $U(1)$ theory for the early universe, in which
the time-temperature relationship $tT^{2} = \Gamma M_{pl}$ is valid,
where we take $\Gamma = O(10^{-1})$ in the GUT era.  If $G$ is
Newton's constant and $\mu$ the classical string tension then,
following \cite{zurek1}, $Mt_{Q}\sim
10^{-1}\lambda^{1/2}(G\mu)^{-1/2}$.  The dimensionless quantity
$G\mu\sim 10^{-6} - 10^{-7}$ is
the small parameter of cosmic string theory.  A value $\lambda\sim
10^{-2}$ gives $Mt_{Q}\sim (Mt_{*})^{a}, a\sim 2$, once factors of
$\pi$, etc.are taken into account, rather than
$Mt_{Q}\sim 1/\lambda$, and the density of  Eq.\ref{nisMf2} may be relevant.

\subsection{Backreaction in QFT}

To improve upon the free-roll result more honestly, but retain the
Gaussian approximation for the field correlation functions, the
best we can do is adopt a mean-field approximation along the lines
of \cite{boyanovsky,LA}, as we did for the CM systems earlier.
As there,  it does have the
correct behaviour of stopping domain growth as the field spreads
to the potential minima.  As before, only the large-$N$ expansion
preserves Goldstone's theorem.

$G(r;t)$ still has the mode decomposition of Eq.\ref{modesum}, but
the modes $\chi^{\pm}_{k}$ now satisfy the equation
\begin{equation}
\Biggl [ \frac{d^2}{dt^2} + {\bf k}^2 + m^2(t) +
\lambda\langle\Phi^{2}({\bf 0})\rangle_{t} \Biggr
]\chi^{\pm}_{k}(t) =0, \label{modeh}
\end{equation}
where we have taken $N=2$. Because $\lambda\phi^{4}$ theory is not
asymptotically free, particularly in the Hartree approximation,
the renormalised $\lambda$ coupling shows a Landau ghost. This
means that the theory can only be taken as a low energy effective
theory.

The end result is\cite{boyanovsky}, on making a single
subtraction at $t=0$, is
\begin{equation}
\Biggl [ \frac{d^2}{dt^2} + {\bf k}^2 + m^2(t) + \lambda\int d \!
\! \! / ^3 p \, C(p) [\chi^{+}_{p}(t)\chi^{-}_{p}(t)-1] \Biggr
]\chi^{\pm}_{k}(t) =0. \label{modeh2}
\end{equation}
which we write as
\begin{equation}
\Biggl [ \frac{d^2}{dt^2} + {\bf k}^2 -\mu^2(t) \Biggr
]\chi_{k}(t)  =0. \label{modemu}
\end{equation}
On keeping just the unstable modes in $\langle\Phi^{2}({\bf 0})\rangle_{t}$
then, as it grows, its contribution
to (\ref{modeh2}) weakens the instabilities, so that only longer
wavelengths become unstable. At $t^{*}$ the instabilities shut
off, by definition, and oscillatory behaviour ensues.
Since the mode with wavenumber $k >0$ stops growing
at time $t^{+}_k <t^{*}$, where $\mu^{2}(t^{+}_{k}) = {\bf k}^{2}$,
the free-roll density at $t^{*}$ must be an overestimate.

An approximation that improves upon the WKB approximation is
\begin{equation}
\chi_{k}(t) \approx \bigg(\frac{\pi M}{2\Omega_{k}(\eta )}\bigg)^{1/2}
\exp\bigg(\int_{0}^{t}dt\,\Omega (t)\bigg) \label{chiKf}
\end{equation}
when $\eta =M(t^{+}_k -t) > 0$ is large, and $\Omega_{k}^{2}(t) = \mu^{2}(t) - {\bf
k}^{2}$. On expanding the
exponent in powers of $k$ and retaining only the quadratic terms
we recover the WKB approximation when $\mu (t)$ is non-zero.

The  result
is that the effect  of the back-reaction is to give a time-delay
$\Delta t$ to $t^*$, corresponding to a decrease in the value
$k_{0}(t)$ at which the power peaks of order
\begin{equation}
\frac{\Delta t}{t^*} = O\bigg(\frac{1}{ln(1/\lambda)}\bigg).
\label{lag}
\end{equation}
The backreaction has little effect for times $t<t^{*}$.
For
$t>t^{*}$ oscillatory modes take over the correlation function
and we expect oscillations in $G(k;t)$.

In practice the backreaction rapidly forces
$\mu^{2}(t)$ towards zero if the coupling is not too
small\cite{boyanovsky}. For couplings that are
not too weak, this requires that we graft
purely oscillatory long wavelength behaviour onto the
non-perturbatively large exponential mode
\begin{equation}
\chi^{+}_{k}(t^{*} )
\approx\alpha_{k}\exp\bigg(\int^{t^{*}}_{0}\,dt'\mu(t')\bigg)
\exp\bigg(-\frac{\sqrt{\tau_Q t^{*}}}{M} k^{2}\bigg)
\label{chipless3}
\end{equation}
The end result is a new power spectrum, obtained by
superimposing oscillatory behaviour onto the old spectrum, frozen at
time $t^{*}$.
As a gross oversimplification, the contribution from the earlier exponential modes
alone can only be to contribute terms something like
\begin{eqnarray}
G(r;t)&\propto& \frac{T}{M|m(t^{*})|}e^{4M(t^{*})^{3/2}/3\sqrt{\tau_Q}}
\int_{|{\bf k}|<M} d \! \! \! / ^3 k
\, e^{i {\bf k} . {\bf x} }
\;e^{-2\sqrt{t^{*}\tau_Q}k^{2}/M}
\nonumber
\\
&\times&\bigg[\cos k(t-t^{*}) +\frac{\Omega (k)- W'(k)}{k}\sin k(t-t^{*})\bigg]^{2}
\label{Wexps}
\end{eqnarray}
to $G$, where $\Omega
= M(t^{*} - t_k)^{1/2}/\tau_Q^{1/2}$ and $W' = 1/4(t^{*}
-t_k )$.  The details are almost irrelevant, since the density of
line zeroes is independent of the normalisation, and only weakly
dependent on the power spectrum.

The $k=0$ mode of Eq.\ref{Wexps} encodes the simple solution
$\chi_{k=0}(t) = a + bt$ when $\mu^2 = 0$. As observed\cite{boyanovsky2} by Boyanovsky
et al. this has built into it the basic causality discussed by
Kibble\cite{kibble3}.  Specifically, for $r,t\rightarrow\infty$, but
$r/2t$ constant $(\neq 1)$,
\begin{equation}
G(r,t)\approx \frac{C}{r}\Theta (2t/r -1).
\end{equation}
It follows directly that this causality, engendered by the
Goldstone particles of the self-consistent theory, has little effect
on the density of line-zeroes that we expect to mature into fully
classical vortices, since that is determined by the behaviour at $r=0$.

Further, for large $t$ the power spectrum effectively has a
$k^{-2}$ behaviour for small $k$, unlike the white noise that
would follow from  Eq.\ref{Wexp}. It has been suggested\cite{andy} that, for
such a spectrum, most, if not all, of the vortices are in loops,
with little or no self-avoiding 'infinite' string (but see
\cite{scherrer}). If there was no infinite string the evolution
of the network could be very different\cite{dani} from that of white noise,
where approximately $75\%$ of the string is 'infinite'\cite{tanmay}.
Although causality due to massless Goldstone modes is unrealistic,
the linking of causal behaviour to the long wavelength spectrum is
general.
It has to be said that this approximation should
not be taken very seriously for large $t$ on different grounds, since we would expect
rescattering to take place at times $\Delta t = O(1/\lambda)$ in a
way that is precluded by the Gaussian approximation.

Returning to our original concerns, if Eq.\ref{ginlim2} is not satisfied, it is
difficult to imagine how clean vortices, or proto-vortices, can
appear later without some additional ingredient.

\section{Conclusions}

We examined the Kibble /Zurek predictions for the onset of phase
transitions and the appearance of defects (in particular, vortices
or global cosmic strings) as a signal of the
symmetry breaking. Our results are in agreement with their prediction Eq.\ref{xiKZ}
as to the magnitude of the correlation length at the time the
transition truly begins, equally true for condensed matter and QFT.

However, this is not simply a measure of the separation of defects
at the time of their appearance.  The time ${\bar t}$ is too early
for the field to have found the true groundstates of the theory. We
believe that time, essentially the spinodal time, is the time
at which proto-vortices can appear, which can later evolve into the
standard classical vortices of the theory.

Even then, they may not appear because of thermal field
fluctuations. In TDLG condensed matter thermal noise is proportional to
temperature. If temperature is {\it fixed}, but {\it not} otherwise,
as in the pressure quenches of
$^{4}He$, this noise can inhibit
the production of vortices, although there are other factors to be
taken into account (such as their decay rate). On the other hand, on
quenching from a high temperature in QFT there are
always thermal fluctuations, and these can also disturb the appearance of
vortices.
 The condition that thermal fluctuations are ignorable at the time
that the field has achieved the true groundstates can be written
\begin{equation}
(\tau_{Q}/\tau_{0})^{\gamma}(1-T_{G}/T_c )<C,
\label{ginlim3}
\end{equation}
where $\gamma = 1$ for condensed matter and $\gamma = 2$ for QFT. $C = O(1)$.

This restores the role of the Ginzburg temperature $T_G$ that the
simple causal arguments overlooked, but does not restore thermal
fluctuations as the
exclusive agent {\it for} vortex production, as happened in early
arguments. Quenches in $^{4}He$ provide the major example for which
Eq.\ref{ginlim3} is not satisfied.

What happens at late time is unclear, although for TDLG numerical
simulations
can be performed (but have yet to address this problem exactly).
On the other hand, not only is the case of a single
self-interacting {\it quantum} scalar field in flat space-time a caricature  of
the early universe, but it is extremely difficult to go beyond the
Gaussian approximation.  To do better requires that we
do differently. There are several possible approaches.  One step is
to take the FRW metric of the early universe seriously, whereby the
dissipation due to the expansion of the universe can change the
situation dramatically\cite{stephens}.  Other approaches are more explicit in their
attempts to trigger decoherence explicitly, as we mentioned earlier. Most simply,
the short wavelength parts of  the field can be treated as an environment to be
integrated over, to give a coarse-grained theory of long-wavelength
modes acting classically in the presence of noise. However, such
noise is more complicated than in TDLG theory, being multiplicative
as well as additive, and coloured\cite{gleiser2,lombardo,muller2}.
This is an area to be pursued elsewhere.

I particularly thank Glykeria Karra and Eleftheria Kavoussanaki,
 with whom much of this work
was done.  I also would like to thank  Tom Kibble
and many of my colleagues working in this area
 for fruitful discussions. This is an area with a substantial, if
scattered, literature and I have aimed to be exemplary, rather than
inclusive. I apologise to any authors who I have not cited explicitly.
This work is the result of a network supported by the European
Science Foundation.

\end{document}